\documentclass[notitlepage,
 aip,
 amsmath,amssymb,
reprint,%
floatfix
]{revtex4-1}

\usepackage[version=3]{mhchem} 
\usepackage{multirow}
\usepackage{threeparttable}
\usepackage[labelsep=period]{caption}
\usepackage{appendix}
\usepackage{braket}
\usepackage{caption}
\usepackage{subcaption}
\usepackage{comment}
\usepackage{color}
\usepackage{graphicx}
\usepackage{dcolumn}
\usepackage{bm}
\usepackage[section]{placeins}
\usepackage{soul}
\usepackage{siunitx}

\usepackage[labelsep=period]{caption}
\renewcommand{\thetable}{\arabic{table}}
\usepackage{multirow}
\usepackage{threeparttable}
\usepackage{eucal}

\begin{document}

\preprint{AIP/123-QED}

\title[]{An Excited-State-Specific \textcolor{black}{Pseudo}projected Coupled-Cluster Theory
}

\author{Harrison Tuckman}
\affiliation{
Department of Chemistry, University of California, Berkeley, California 94720, USA 
}

\author{Eric Neuscamman}
\email{eneuscamman@berkeley.edu}
\affiliation{
Department of Chemistry, University of California, Berkeley, California 94720, USA 
}
\affiliation{Chemical Sciences Division, Lawrence Berkeley National Laboratory, Berkeley, CA, 94720, USA}

\date{\today}

\begin{abstract}
  We present an excited-state-specific coupled-cluster approach in
which both the molecular orbitals and cluster amplitudes are
optimized for an individual excited state.
The theory is formulated via a \textcolor{black}{pseudo}projection of
the traditional coupled-cluster wavefunction that allows 
correlation effects to be introduced atop
an excited state mean field starting point.
The approach shares much in common with ground state CCSD,
including size \textcolor{black}{extensivity} and an $N^6$ cost scaling.
Preliminary numerical tests show that, \textcolor{black}{when augmented with $N^5$-cost perturbative corrections for key terms,} the method can improve over
excited-state-specific second order perturbation theory
in valence, charge transfer, and Rydberg states.
\end{abstract}

\maketitle

\section{Introduction}

Mean field methods such as Hartree-Fock (HF) theory
\cite{szabo2012modern}
are useful for creating
qualitatively correct depictions of a molecules' electronic structure,
but the finer details of electron correlation must also be addressed
in order to achieve quantitative energy predictions.
When working with ground state molecules near their equilibrium geometries,
wavefunction methods traditionally capture this correlation
by hierarchically applying perturbation based and/or
coupled-cluster (CC) based corrections to the HF wavefunction.
\cite{moller1934note,vcivzek1966correlation,vcivzek1971correlation,bartlett2007coupled}
An especially noteworthy part of this hierarchy is coupled-cluster with singles,
doubles, and perturbative triples (CCSD(T)),\cite{raghavachari1989fifth,watts1993coupled}
which is often referred to as the ``gold standard'' thanks to its ability
to deliver sub-$\mathrm{k_B}$T accuracy at polynomial cost in a variety of chemically
important settings.
\cite{thomas1993balance,helgaker1997prediction,bak2001accurate}
In practice, of course, wavefunction methods are often eschewed in favor
of the more computationally efficient density functional theory (DFT), \cite{hohenberg1964inhomogeneous,kohn1965self,parr1980density,piela2013ideas}  
but, one way or another, accurate chemical predictions rely on correctly
accounting for the effects of electron correlation.

Achieving quantitative predictions in electronically excited states
becomes more complicated, as methods must now contend with
open-shell spin-recombination, post-excitation orbital relaxations,
and the effects of these on the details of electron correlation.
In practice, many of the most widely used 
excited state methods such as time-dependent DFT (TD-DFT)\cite{runge1984density,burke2005time,casida2012progress}, and linear response CC (LR-CC)\cite{monkhorst1977calculation,koch1994calculation} approach excited states by way of linear response theory. \textcolor{black}{Equation-of-motion CC (EOM-CC),\cite{rowe1968equations,stanton1993equation,krylov2008equation} which is often motivated from a lens not rooted explicitly in linear response theory, ends up bearing many similarities to LR-CC, including matching excitation energies.\cite{koch1994calculation,caricato2009difference} Though LR-CC and EOM-CC differ for some other properties such as transition moments, the primary focus of this paper is excitation energies, and so we will simply refer to both these theories as linear response approaches. While all of these linear response CC approaches} capture open-shell singlet character
and have a long record of success, they are not without their difficulties.
Typically, the linear response approximation requires the assumption that key aspects
of the electronic structure are shared between ground and excited states.
For example, \textcolor{black}{much of the correlation treatment in EOM-CC and LR-CC comes from the shared exponential operator.}
Likewise, the widely used adiabatic approach to TD-DFT retains the same orbital
shapes for all spectator electrons, regardless of how substantially an excitation
has changed the electrostatic environment that these electrons sit in.
These assumptions can be problematic in situations involving
large and thus nonlinear post-excitation relaxations,
such as many charge transfer, Rydberg, core, and double excitations.
\cite{tozer1999does,sobolewski2003ab,dreuw2003long,dreuw2004failure,mester2022charge,izsak2020single,kozma2020new,tozer1998improving,casida1998molecular,casida2000asymptotic,tozer2003importance,tozer2000determination,maitra2004double,cave2004dressed,levine2006conical}
Linear response methods can also struggle in
situations where the ground state is not well described by mean field theory,
as they can inherit the ground state difficulties even if the structure of the
excited state is simple.
While there are many different corrections \textcolor{black}{and augmentations} that can be added to
these methods in order to address specific shortcomings --
e.g.\ long range corrections in TD-DFT \cite{tawada2004long,yanai2004new} \textcolor{black}{or charge transfer separability corrections in EOM-CCx \cite{musial2011charge}} --
the various challenges facing linear response methods have prevented the
community from coalescing around a unifying ``gold standard''
and have left excited state modeling at a distinct disadvantage relative to ground states.

Motivated in part by the limitations of linear response, there has been
much recent work on excited-state-specific methods.
Whether studying single determinant wavefunctions,
\cite{ziegler1977calculation,kowalczyk2011assessment,gilbert2008self,besley2009self,barca2018simple}
configuration interaction (CI) wavefunctions, \cite{kossoski2023state}
CC wavefunctions, \cite{mayhall2010multiple,lee2019excited,folkestad2023entanglement,kossoski2021excited,marie2021variational,arias2022accurate,rishi2023dark}
or DFT functionals, \cite{kowalczyk2013excitation,hait2020excited,hait2021orbital}
the theme of locating higher energy, state-specific excited state and open-shell roots
is becoming increasingly prevalent across the field.
Recognizing that, in the ground state, CCSD is a crucial stepping stone
towards CCSD(T) as well as a useful method in its own right,
our focus here is to construct a CCSD-like excited-state-specific
CC theory atop an excited state mean field (ESMF)
starting point \cite{shea2018communication,shea2020generalized,hardikar2020self}
that, like HF in the ground state, has already taken care of state-specific orbital relaxations.
Besides, with an $N^5$ cost ESMF-based second order perturbation theory (ESMP2)
already established, \cite{clune2020n5}
the longstanding pattern in ground state correlation methods makes it seem
natural to expect an $N^6$ cost CC method to provide the next step
in a hierarchy of excited-state-specific correlation methods.

Creating this CC method poses immediate challenges, however, due to the
inherently multireference nature of excited states, and so it is prudent to consider
the pros and cons of existing approaches to multireference CC theory.
Such methods typically fall into one of three categories:
the Jeziorski-Monkhort (JM) based methods,
the internal contraction (ic) based methods,
and the single-reference based methods.
\cite{kohn2013state,lischka2018multireference,lyakh2012multireference}
While research continues into notable multi-reference coupled-cluster methods which make use of the Jeziorski-Monkhorst ansatz such as State-Universal (SU)-MRCC,\cite{jeziorski1981coupled,kucharski1991hilbert,piecuch2002state} Brilluoin-Wigner (BW)-MRCC,\cite{hubavc1994size,pittner1999assessment} single root (sr)-MRCC,\cite{mahapatra2010potential,mahapatra2011evaluation} Mukherjee (Mk)-MRCC,\cite{mahapatra1998state,mahapatra1999size} MRexpT,\cite{hanrath2005exponential} and two-determinant (TD)-CC,\cite{balkova1992coupled,szalay1994analytic,lutz2018reference} as well as methods which make use of internally contracted schemes such as multiple flavors of ic-MRCC,\cite{hanauer2011pilot,evangelista2011orbital,hanauer2012communication} partially internally contracted (pIC)-MRCC,\cite{datta2011state} and block correlated CC,\cite{li2004block} these methods are not without their difficulties.
In general, the state-universal JM methods can face numerical instabilities due to the intruder state
problem, \cite{lyakh2012multireference} while the state-specific approaches have an inherent mismatch
between the number of cluster amplitudes and residual equations.
\cite{kohn2013state,lischka2018multireference}
Furthermore, both of these methods lack an invariance to orbital rotations within the active space,\cite{evangelista2010insights,kong2010orbital} making the final energy dependent on the choice of orbitals.
Meanwhile, the internally contracted methods must contend with both overlap matrices that
can introduce their own numerical instabilities and, in some cases, approximate truncations
of nonterminating BCH expansions.
\cite{kohn2013state,lischka2018multireference,lyakh2012multireference,neuscamman2010strongly,neuscamman2010review}

To avoid these challenges and maintain as much of the convenience of single-reference CC
as possible, we will therefore attempt to tackle multi-reference character via an approach
similar to the ``little t'' and ``little tq'' methods,
\cite{oliphant1991multireference,piecuch1999coupled,shen2012combining,bauman2017combining,piecuch1993state,adamowicz2000new,lyakh2008generalization}
which rely on a selective inclusion of
only a subset of triples and/or quadruples.
The subset in question has often been identified by defining
an active space, but more recent work has also explored more general approaches.
\cite{deustua2017converging,deustua2018communication}
In a manner similar to the $N^5$ variant of ESMP2, \cite{clune2020n5}
we propose to use the ESMF starting point to automatically identify a subset of
triples to include, thus limiting the number of triples in a way that makes an $N^6$
scaling possible and avoids user-defined active spaces.
With the introduction of a projection\textcolor{black}{-like} operator that suppresses the
Aufbau configuration while maintaining size \textcolor{black}{extensivity and intensivity}, and with the help
of a substantial dose of automated algebra,
we find that this approach leads to a fully excited-state-specific, CCSD-like
theory for adding CC-style correlation to an ESMF starting point.
\section{Theory}

\subsection{Conventional Ground State Coupled-Cluster}

While there exist a plethora of resources illustrating ground state coupled-cluster theory in extensive detail,\cite{crawford2007introduction,helgaker2013molecular,bartlett2007coupled} we wish to provide a quick summary here so that we may draw parallels between our excited state method and the ground state theory. Although CC theory is at its heart an approximation to the many-body energy, it can be motivated by a wavefunction that is defined by operating the exponential of a cluster operator, $\hat{T}$, on a reference, $\ket{\phi _0}$, which is most often the Hartree-Fock determinant,

\begin{equation}
  \ket{\psi}=e^{\hat{T}}\ket{\phi _0}=\left(1+\hat{T}+\frac{1}{2}\hat{T}^2+\dots\right)\ket{\phi _0}\label{eqn: exponential}.  
\end{equation}

One of the challenges in applying this exponential operator is that obtaining a variational solution for the energy necessitates either exponentially scaling computational cost or an approximate truncation of the exponential expansion. Therefore, what is often done in practice is to utilize a projective approach where cluster amplitudes are determined by satisfying the eigenvector condition for the similarity transformed Hamiltonian in a chemically relevant subspace. For example, in CC with single and double excitations (CCSD), $\hat{T}$ is truncated to include only $\hat{T}_1$ and $\hat{T}_2$ \textcolor{black}{while} projections are taken against the reference determinant, $\ket{\phi _0}$, and all singly and doubly excited determinants, $\ket{\phi _i ^a}$ and $\ket{\phi _{ij} ^{ab}}$, in order to define an approximate (and non-variational) energy equation and a set of amplitude equations that can be solved to determine $\hat{T}_1$ and $\hat{T}_2$.\cite{crawford2007introduction,helgaker2013molecular,bartlett2007coupled}

\begin{align}
    E_{CCSD}&=\bra{\phi _0}e^{-(\hat{T}_1 + \hat{T}_2)}\hat{H}e^{\hat{T}_1 + \hat{T}_2}\ket{\phi _0}, \label{eqn: groundE}\\
    0&=\bra{\phi _i ^a}e^{-(\hat{T}_1 + \hat{T}_2)}\hat{H}e^{\hat{T}_1 + \hat{T}_2}\ket{\phi _0}, \label{eqn: groundS}\\
    0&=\bra{\phi _{ij} ^{ab}}e^{-(\hat{T}_1 + \hat{T}_2)}\hat{H}e^{\hat{T}_1 + \hat{T}_2}\ket{\phi _0} \label{eqn: groundD}.
\end{align}

\noindent Here and in the rest of this paper, $i,j,k,\dots$ represent occupied spin orbitals while $a,b,c,\dots$ represent virtual spin orbitals relative to the reference. Solving the above equations produces the energy and cluster amplitudes, which can in turn be used to evaluate other properties.\cite{bartlett2007coupled} 

\subsection{A Suitable Reference and the Usefulness of a \textcolor{black}{Pseudo}projector}

\begin{table}[htb]
    \centering
    \begin{tabular}{c c c c}
         Wavefunction & Description & Equation & $\quad$\textcolor{black}{MO Basis}$\quad$ \\ \hline
         $\ket{\phi _0}$ & RHF & --- & \textcolor{black}{RHF} \\
         \multirow[t]{ 2}{*}{$\ket{\phi ^* _0}$} & ESMF Aufbau & \multirow[t]{2}{*}{\ref{eqn: ESMF}} & \multirow[t]{2}{*}{ESMF TOP}\\
         &Formal Reference&&\\
         $\ket{\psi  _{ESMF}}$ & ESMF & \ref{eqn: ESMF} & \textcolor{black}{ESMF TOP}\\
         \multirow[t]{ 2}{*}{$\ket{\psi _0}$} & Truncated ESMF & \multirow[t]{2}{*}{\ref{eqn: truncated esmf}} & \multirow[t]{2}{*}{ESMF TOP}\\
         &Reference&&\\
         $\ket{\Psi }$ & ESPCC & \ref{eqn: my wavefunction} & \textcolor{black}{ESMF TOP}\\
    \end{tabular}
    \caption{A brief summary of the wavefunctions discussed in this paper, a short description of what they represent, the relevant equation where they are defined, and the MO basis they are constructed in.}
    \label{tab: wavefunctions}
\end{table}

As is the case for the ground state, choice of a reference is the first step in our CC method. For studying electronically excited states, we elect to use the Excited State Mean Field (ESMF) method,  \cite{shea2018communication,shea2020generalized,hardikar2020self} which is an excited state analog to the mean field HF reference, as a starting point. In a spin orbital basis, the ESMF wavefunction can, if we neglect the possibility of the Aufbau determinant participating in the excited state, be expressed as

\begin{equation}
    \ket{\psi _{ESMF}}=e^{\hat{X}}\left( \sum _{ia} c_{ia}\hat{a}_a ^\dagger \hat{a}_i \ket{\phi _0}\right)=\sum _{ia} c_{ia}\hat{a}_a ^\dagger \hat{a}_i \ket{\phi^* _0}. \label{eqn: ESMF}
\end{equation}

\noindent Here $e^{\hat{X}}$ is an orbital rotation operator and $\hat{a}_a ^\dagger$ and $\hat{a}_i$ are second quantization creation and destruction operators respectively. \textcolor{black}{To arrive at the expression on the right of equation \ref{eqn: ESMF}, the orbital rotation operator is acted on the original canonical orbital basis in order to result in a new basis whose orbitals are optimized for the specific excited state of interest. We denote this new basis with an asterisk, such that the Aufbau determinant in this ESMF orbital basis is denoted as $\ket{\phi ^* _0}$.} Immediately it is evident that our reference is a multi-determinant wavefunction, which can complicate CC methods.\cite{lyakh2012multireference,kohn2013state} We can mitigate the extent of the multireference nature of our wavefunction by performing a singular value decomposition (SVD) on the CI coefficients of the ESMF wavefunction and transforming our basis to a transition orbital pair (TOP) basis. \cite{clune2020n5} In the TOP basis --- \textcolor{black}{which shares many similarities with the natural transition orbital basis \cite{martin2003natural} and has been employed by methods such as EOM CC to reduce computational scaling  \cite{park2018low}} --- each occupied index becomes paired with a virtual index, thereby reducing the number of determinants in our ESMF wavefunction. We can further reduce the number of determinants by truncating the ESMF wavefunction after a specified threshold. Although we will discuss generalizations below, in this initial investigation we choose to employ the most aggressive truncation scheme possible by truncating after the first singular value. For now, \textcolor{black}{this results in our} (renormalized) truncated ESMF \textcolor{black}{reference} wavefunction, $\ket{\psi _0}$,

\begin{equation}
    \ket{\psi _0}=\frac{1}{\sqrt{2}}\left(\ket{{\phi ^*} _{h} ^{p}}+\ket{{\phi ^*} _{\bar{h}} ^{\bar{p}}}\right). \label{eqn: truncated esmf}
\end{equation}

\noindent Here any index with a bar denotes the opposite electron spin as compared to the index without a bar, and $h$ and $p$ are the special paired indices which correspond to the orbital that is primarily excited out of in the occupied space and the orbital that is primarily excited into for a particular excited state. Because these orbitals are so important for describing the excited state of interest, we will henceforth refer to them as the hole orbitals and the particle orbitals respectively. It is worth pointing out that our truncated ESMF \textcolor{black}{reference}, $\ket{\psi _0}$, bears many similarities to the reference in TD-CC,\cite{balkova1992coupled,szalay1994analytic,lutz2018reference} as both wavefunctions represent the simplest possible state capable of describing open shell singlet excited states. 
As we will explain presently, however, our formulation differs significantly from TD-CC due to its ``little t''
structure and its projection\textcolor{black}{-like} operator.

Following the general strategy of CCSDt \cite{piecuch1999coupled,oliphant1991multireference}
but using the ESMF TOP orbital basis,
we apply $\mathrm{exp}(\hat{T})$ 
to the Aufbau determinant, $\ket{\phi ^* _0}$, \textcolor{black}{which will serve as our ``formal reference''. Through a pseudoprojector and a careful initialization of our $\hat{T}$ operator, our formalism transforms the closed-shell, single-determinant $\ket{\phi ^* _0}$ formal reference into the truncated ESMF reference $\ket{\psi _0}$ in a way that avoids many of the challenges typically associated with a multi-determinant reference. Of course,} applying the exponential ansatz directly to the Aufbau determinant creates an immediate issue for excited states. Looking at the Taylor expansion of the exponential ansatz in equation \ref{eqn: exponential}, we see that the wavefunction would have a large contribution from the Aufbau determinant regardless of our initialization of $\hat{T}$, which is neither typical in excited states nor present in our desired ESMF \textcolor{black}{reference}.
We address this issue by introducing a \textcolor{black}{pseudo}projection operator $\hat{P}$ into
the definition of our excited state \textcolor{black}{pseudo}projected coupled-cluster (ESPCC) wavefunction.
\begin{align}
    \ket{\Psi}&=\hat{P}e^{\hat{T}}\ket{\phi ^* _0} \label{eqn: my wavefunction}\\ 
    \hat{P}&=2-\hat{a}_{h}^{\dagger}\hat{a}_{h}-\hat{a}_{\bar{h}}^{\dagger}\hat{a}_{\bar{h}}
    \label{eqn: projector}
\end{align}
\noindent Here $\hat{a}_{h} ^\dagger$/$\hat{a}_{\bar{h}} ^\dagger$ and $\hat{a} _{h}$/$\hat{a} _{\bar{h}}$ are the creation and destruction operators for the hole orbitals that are excited out of in the ESMF \textcolor{black}{reference}, and the 2 comes from the total number of these orbitals (one alpha and one beta). $\hat{P}$ deletes determinants in which the hole orbital is doubly occupied, and so in particular deletes the Aufbau determinant.

\textcolor{black}{Like a true projector, the pseudoprojector $\hat{P}$
deletes certain pieces of the wavefunction like the Aufbau determinant, but, unlike a true projector, it modifies the coefficients on some of the pieces that are not deleted.}
\textcolor{black}{Note that we avoid using the seemingly more straightforward true projector  $\hat{P}_A=1-\ket{\phi ^* _0}\bra{\phi ^* _0}$ whose only effect is to delete the Aufbau contribution, because it breaks size intensivity. As discussed in Section \ref{sec:sc} below, the pseudoprojector in equation \ref{eqn: projector} maintains both size extensivity and size intensivity. We also} note that by deviating slightly from a rigid value of 2 in equation \ref{eqn: projector}, we could introduce the flexibility to allow some Aufbau contribution, which does occur for excited states in the same symmetry representation as the ground state. One could also consider multi-orbital generalizations of $\hat{P}$ that could work with a less truncated ESMF \textcolor{black}{reference}, but for simplicity in this initial study we stick to the \textcolor{black}{pseudo}projector in equation \ref{eqn: projector}. It's worth emphasizing that the details of the hole orbitals in the \textcolor{black}{pseudo}projection operator are completely specified by the ESMF procedure. To help organize the wavefunctions presented so far, we have compiled them in Table \ref{tab: wavefunctions}.


\subsection{The Cluster Operator}

\label{sec:cluster-op}

\begin{figure*}[t]
    \centering
    \includegraphics[width = 0.95\linewidth]{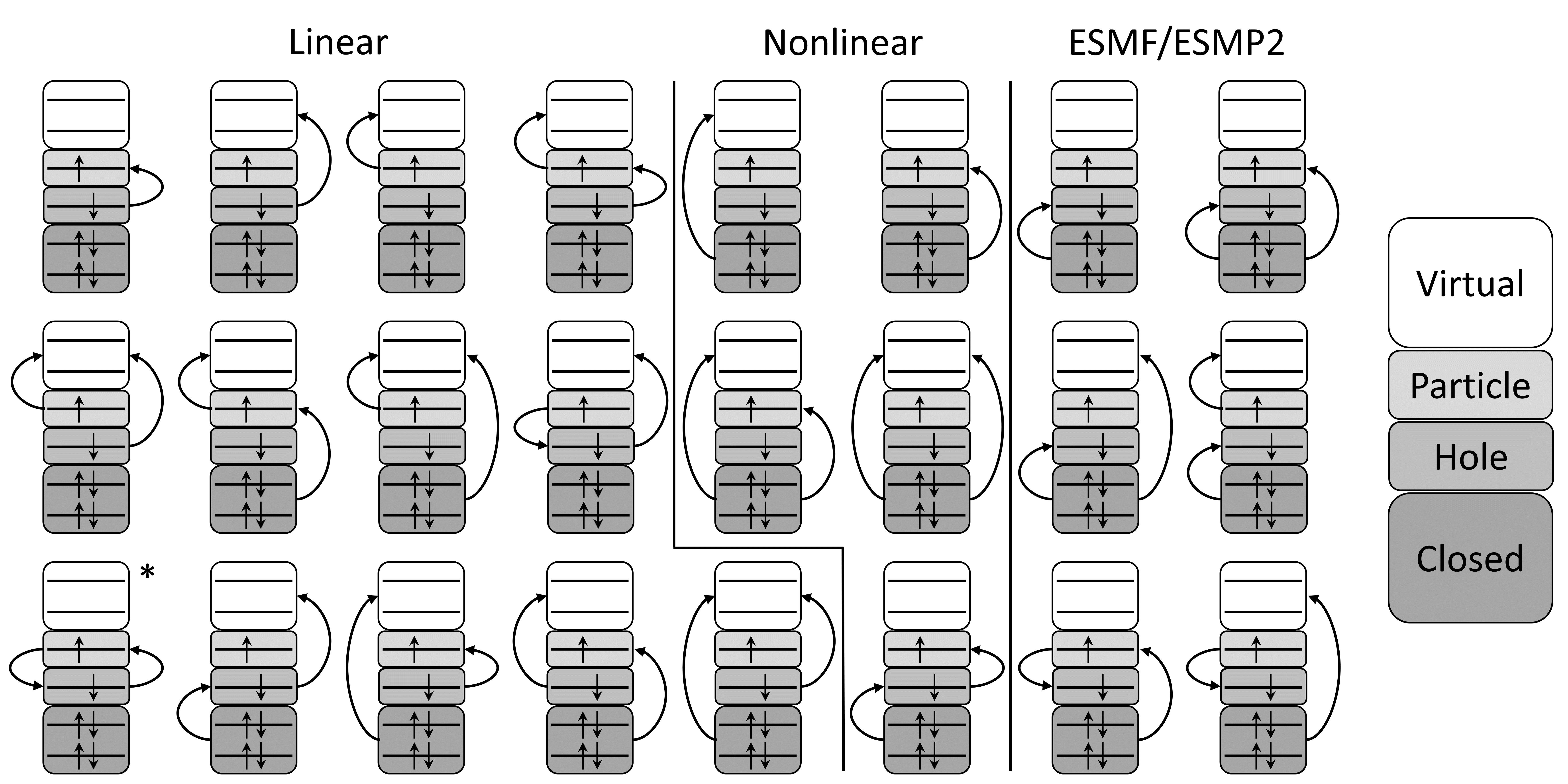}
    \caption{\small Presented here are all determinants within two excitations of one of the two determinants comprising the truncated ESMF \textcolor{black}{reference}.
    Note that the Aufbau determinant is intentionally missing, as it is
    removed by our \textcolor{black}{pseudo}projector.
    The orbital space can be divided into four distinct regions:
    the closed space, hole space, particle space, and virtual space.
    The closed and hole spaces together form the set of orbitals which are occupied in the Aufbau determinant, while the particle and virtual spaces combine to form the set of unoccupied orbitals for the Aufbau determinant. Furthermore, the hole and particle TOP orbitals comprise the set of orbitals which are primarily excited out of and into respectively for the excited state of interest and are determined by an SVD of the \textcolor{black}{original} ESMF starting point. The entire set of determinants is also partitioned according to where they appear in the current theory. Those determinants under the linear heading either contain a hole in the hole space but not a TOP or contain multiple holes in the hole space, and thus are treated primarily by excitation operators appearing at linear order in the Taylor expansion of the exponential ansatz. In contrast, those under the nonlinear heading contain a single TOP with no additional holes in the hole space, and therefore see large contributions from the quadratic order $\hat{T}^2$ terms. Finally, those under the ESMF/ESMP2 heading are missed in the current implementation because they contain completely filled hole orbitals and thus are destroyed by our \textcolor{black}{pseudo}projector, though their contribution is later estimated and corrected for via both the ESMF and ESMP2 theories. The determinant marked by an `*' corresponds to the spin flipped TOP, which while included in the theory has its amplitude fixed to achieve intermediate normalization.}
    \label{fig:determinants}
\end{figure*}

As is the case in other single-reference-based CC methods aimed at strongly multireference situations, \cite{oliphant1991multireference,piecuch1999coupled,shen2012combining,bauman2017combining} a choice must be made as to which excitation operators are included in the wavefunction. In order to produce an analog to ground state CCSD, we wish to include the excitations necessary to produce all determinants that are a single or double excitation away from the truncated ESMF \textcolor{black}{reference}, as illustrated in Figure \ref{fig:determinants}. To do so, we require a subset of triple excitations relative to the Aufbau determinant, which leads to the following form for our cluster operator,

\begin{equation}
    \hat{T}=\hat{T}_1+\hat{T}_{2}+\hat{T}_{\Tilde{3}},
\end{equation}

\noindent where $\hat{T}_{\Tilde{3}}$ represents only a small subset of all possible triples, specifically those which include a hole-particle TOP. Note that, despite the fact that not all double excitations off the Aufbau determinant, $\ket{\phi ^* _0}$, are within the first order interacting space of our reference, $\ket{\psi _0}$, all of the double excitation operators must be included in order to ensure size \textcolor{black}{intensivity}, as will be illustrated in Section \ref{sec:sc}.
As stated above, we will initialize our CC ansatz such that it reconstructs the truncated ESMF wavefunction in equation \ref{eqn: truncated esmf}.
\textcolor{black}{By writing the relevant ESMF CI coefficients in CC form, i.e.\ by setting $t_h^p=t_{\bar{h}}^{\bar{p}}=\frac{1}{\sqrt{2}}$ and $t_{h \bar{h}}^{p \bar{p}}=-\frac{1}{2}$, we can exactly reconstruct the truncated ESMF state as the initial guess for our CC wavefunction $\ket{\Psi}$.}

The introduction of the \textcolor{black}{pseudo}projector in equation \ref{eqn: my wavefunction} adds an extra layer of complexity that is not present in the ground state theory. In the Taylor series expansion of our wavefunction, components of $\hat{T}$ that contain a hole index survive the \textcolor{black}{pseudo}projector and appear at linear order, whereas the linear order appearances of components without the special hole index are eliminated by the \textcolor{black}{pseudo}projector. Thus, these components only contribute to $\ket{\Psi}$ via $\hat{T}^2$ or higher order terms in which they appear alongside a component with the special hole index.
We expect that $t_h^p$ and $t_{\bar{h}}^{\bar{p}}$
will be the largest components of $\hat{T}_1$, so the largest contributions
from excitation operators that do not contain a hole index will appear
inside $\hat{T}^2$, where they will show up in cross terms with
$t_h^p$ and $t_{\bar{h}}^{\bar{p}}$.
So, for a cluster amplitude such as $t _{jk} ^{bc}$ that contains no hole index,
the primary contribution will be to the part of Hilbert space projected to by
$\bra{{\phi ^*} _{hjk}^{pbc}}$ and $\bra{{\phi ^*} _{\bar{h}jk}^{\bar{p}bc}}$.
In stark contrast to ground state CCSD, the projection of Hilbert space usually associated with this amplitude, $\bra{{\phi ^*} _{jk}^{bc}}$, is actually \textcolor{black}{relatively} unimportant, as it is more than two excitations away from the truncated ESMF \textcolor{black}{reference}. In the amplitude equations below, we will project with $\bra{{\phi ^*} _{hjk}^{pbc}}+\bra{{\phi ^*} _{\bar{h}jk}^{\bar{p}bc}}$ and then ensure that the corresponding projection, $\bra{{\phi ^*} _{hjk}^{pbc}}-\bra{{\phi ^*} _{\bar{h}jk}^{\bar{p}bc}}$, is satisfied by enforcing $t _{hjk} ^{pbc}=-t _{\bar{h}jk} ^{\bar{p}bc}$, which has the added benefit of avoiding any potential redundancy in the amplitudes. It is worth noting that, unlike the sufficiency conditions often introduced in Jeziorski-Monkhort based MRCC methods, the amplitude equations resulting from this system of linear combinations of projections are uniquely satisfied when the individual projections are satisfied, and thus satisfy the proper residual problem.\cite{kohn2013state,lischka2018multireference,lyakh2012multireference} Furthermore, while it might be tempting to instead simply remove any parts of $\hat{T}$ that only contribute through $\hat{T}^2$ and higher terms, these components are critically important for maintaining size \textcolor{black}{intensivity}, as will be detailed in Section 2.6.  

Lastly, we set the theory's definition of intermediate normalization by fixing at $\frac{1}{\sqrt{2}}$ the $t_h^p$ and $t_{\bar{h}}^{\bar{p}}$ amplitudes used to construct the \textcolor{black}{truncated} ESMF \textcolor{black}{reference}. This is analogous to how the Aufbau coefficient is always 1 in the expanded CCSD wavefunction. Here, our choice of intermediate normalization ensures a unique solution to the forthcoming amplitude equations and also creates \textcolor{black}{the following} parallel to the ground state theory,
\begin{gather}
    \braket{\psi _0 | \Psi}=\bra{\psi _0}\hat{P}e^{\hat{T}}\ket{\phi ^* _0}=1, \label{eqn: intermediate normalization}
\end{gather}
which matches the CCSD expression,
\begin{gather}
    \braket{\phi _0 | \psi _{CCSD}}=\bra{\phi _0}e^{\hat{T}}\ket{\phi _0}=1.
\end{gather}

\subsection{The Energy and Amplitude Equations}

With an excited-state-appropriate CC wavefunction to get us started, we now invoke the usual CC procedure to produce an approximate, non-variational energy equation as well as a set of amplitude equations. To obtain the energy equation, we project against the truncated ESMF reference, $\ket{\psi _{0}}$:

\begin{equation}
    E\equiv \frac{\bra{\psi _{0}}e^{-\hat{T}}\hat{H}\ket{\Psi }}{\bra{\psi _{0}}e^{-\hat{T}}\ket{\Psi}} =\frac{\bra{\psi _{0}}e^{-\hat{T}}\hat{H}\hat{P}e^{\hat{T}}\ket{\phi ^* _0}}{\bra{\psi _{0}}e^{-\hat{T}}\hat{P}e^{\hat{T}}\ket{\phi ^* _0}}. \label{eqn: energy}
\end{equation}

With the intermediate normalization introduced in the previous section, the denominator in this expression evaluates to 1. Similarly, the amplitude equations are defined by taking projections against all possible excited determinants no more than two excitations from the truncated ESMF \textcolor{black}{reference}, as are illustrated in Figure \ref{fig:determinants}:

\begin{align}
    \bra{{\phi ^*} _i ^a}e^{-\hat{T}}\left(\hat{H}-E\right)\hat{P}e^{\hat{T}}\ket{\phi ^* _0}&=0, \label{eqn: single}\\
    \bra{{\phi ^*} _{ij} ^{ab}}e^{-\hat{T}}\left(\hat{H}-E\right)\hat{P}e^{\hat{T}}\ket{\phi ^* _0}&=0, \label{eqn: double} \\
    \bra{{\phi ^*} _{hjk} ^{pbc}}e^{-\hat{T}}\left(\hat{H}-E\right)\hat{P}e^{\hat{T}}\ket{\phi ^* _0}&=0. \label{eqn: triple}
\end{align}

\noindent Note that we do not have amplitude equations for all singles, doubles, and especially triples, only those in the ``linear'' and ``nonlinear'' categories in Figure \ref{fig:determinants}. While satisfying the remaining singles and doubles amplitude equations under the ``ESMF/ESMP2'' category of Figure \ref{fig:determinants} would make for an ideal parallel to ground state CCSD, with the simple \textcolor{black}{pseudo}projector scheme that we use here, our wavefunction does not have the flexibility to access these regions of Hilbert space, although we anticipate that more general choices of the \textcolor{black}{pseudo}projector may resolve this issue in the future. For now, the energetic contributions of these missing terms will be estimated and corrected for by taking the corresponding terms' energy contributions from untruncated ESMF and ESMP2. Finally, with all singles and doubles present in $\hat{T}$ despite the fact that
we only include singles and doubles bra states within
two excitations of our \textcolor{black}{truncated ESMF reference} in
equations \ref{eqn: single} and \ref{eqn: double},
it may seem at first glance that we have a mismatch between the number
of amplitudes and amplitude equations.
However, as will be illustrated in the following section, the seemingly extra singles and doubles amplitudes go along with the single-TOP-containing doubles and triples amplitude equations, respectively, leaving us with the exact same number of amplitude equations as we have amplitudes to solve for. 

\subsection{Solving the Amplitude Equations}

As mentioned in Section 2.3, there exist two distinct groups of excitation operators: those which contain a hole orbital as one of their indices and those that do not. Those which do survive the \textcolor{black}{pseudo}projection operator and contribute at linear order in the Taylor expansion of the wavefunction, while those which do not are destroyed by the \textcolor{black}{pseudo}projector and therefore first appear at quadratic order where they must be paired with an excitation that does contain a hole index. As a result, we find it effective to use different approaches for these two distinct classes of amplitudes in the iterative solution of the amplitude equations.

To solve the amplitude equations, we make use of a quasi-Newton iterative scheme reminiscent of that often used in ground state CC.\cite{helgaker2013molecular} To calculate the approximate Jacobian, we begin by approximating the Hamiltonian as only the diagonal portion of the Fock operator constructed from the ESMF density matrix. It is worth noting that during the ESMF optimization the orbitals were rotated such that they are no longer in a canonical regime.
\cite{shea2018communication,shea2020generalized,hardikar2020self}
However, we expect that the Fock matrix should remain \textcolor{black}{mostly} diagonally dominant, so we ignore the
off-diagonal terms.
\textcolor{black}{While this approximation was successful for achieving tight convergence in the states tested within this paper, it is worth noting that more robust optimization algorithms whose approximate Jacobians includes off diagonal elements are possible. In the present algorithm, for}
a given cluster amplitude $t_\mu$ containing either a hole index without a TOP or multiple hole indices, we obtain a diagonal approximate Jacobian which leads to an update scheme very similar to that in the ground state theory:
\begin{equation}
    t^{new} _\mu=t^{old} _\mu -\frac{\bra{\phi ^* _\mu}e^{-\hat{T}}(\hat{H}-E)e^{\hat{T}}\ket{\phi ^* _0}}{N_\mu \left(\Delta _\mu + (E - E_0)\right) }. \label{eqn: primary update}
\end{equation}
Here $\bra{\phi ^* _\mu}$ corresponds to the Hilbert subspace resulting from the action of $\hat{T}_\mu$ on $\ket{\phi _0 ^*}$, $\Delta _\mu$ is the orbital energy difference (using the diagonal elements of our Fock operator) corresponding to the $\hat{T}_\mu$ excitation, $N_\mu$ is the number of hole indices in the cluster amplitude, and $E _0=\bra{\phi _0 ^*} \hat{H} \ket{\phi _0 ^*}$.

For the cluster amplitudes that lack a hole index,
the update is a little less conventional and is paired
with the update for the related one-higher-excitation
amplitude that contains the hole-particle TOP.
As discussed in Section \ref{sec:cluster-op},
the most important contribution from a no-hole cluster
amplitude $t_\nu$ occurs in the terms inside $\hat{T}^2$
in which it partners with $t_h^p$ or $t_{\bar{h}}^{\bar{p}}$
in an overall excitation that is one higher than
$t_\nu$ itself.
Thus, we update $t_\nu$ in conjunction with 
the amplitudes $t_\omega$ and $t _{\bar{\omega}}$,
which are like $t_\nu$ but contain an additional
TOP excitation.
For example, if $t_\nu$ is $t_{jk}^{bc}$, then $t_\omega$ is $t_{hjk}^{pbc}$
and $t_{\bar{\omega}}$ is $t_{\bar{h}ij}^{\bar{p}ab}$.
\begin{align}
    t^{new} _\nu&=t^{old} _\nu -\frac{\left(\bra{\phi ^* _{\omega}}+\bra{\phi ^* _{\bar{\omega}}}\right)e^{-\hat{T}}(\hat{H}-E)e^{\hat{T}}\ket{\phi ^* _0}}{(t_h ^p + t_{\bar{h}} ^{\bar{p}}) \Delta _\nu }, \label{eqn: phase I}\\
    t^{new} _{\omega} &=t^{old} _{\omega} -\frac{\left(\bra{\phi ^* _{\omega}}-\bra{\phi ^* _{\bar{\omega}}}\right)e^{-\hat{T}}(\hat{H}-E)e^{\hat{T}}\ket{\phi ^* _0}}{2 \left(\Delta _{\omega} + (E - E_0)\right)}, \label{eqn: phase II}\\
    t^{new} _{\bar{\omega}} &= -t^{new} _{\omega}
    \label{eqn: phase IIb}
\end{align}
Crucially, only two of these three amplitudes are
independent variables
($t_{\bar{\omega}}$ being defined as the negative
of $t_\omega$ as discussed in Section \ref{sec:cluster-op}),
which is why we are able to arrive
at this style of iterative update from just two amplitude
equations (the ones involving
$\bra{\phi ^* _{\omega}}$ and
$\bra{\phi ^* _{\bar{\omega}}}$) without encountering
a linear dependency.
Also note that, in our present formalism, $t_h^p$ and $t_{\bar{h}}^{\bar{p}}$
are fixed at $1/\sqrt{2}$ by intermediate normalization.

In addition to the above approach to iterative amplitude updates, we choose to employ a two phase optimization procedure. While a two phase procedure is not strictly necessary, we find that it assists in achieving rapid and tight convergence of the residual equations. In the first phase, the two ``higher order'' excitation amplitudes $t _\omega$ and $t _{\bar{\omega}}$ are held fixed at zero, leaving only the amplitude associated with equation \ref{eqn: phase I}, $t_\nu$, to change. After this subset of residual equations is sufficiently converged, the remaining amplitudes associated with equations \ref{eqn: phase II} and \ref{eqn: phase IIb} are introduced and all the residual equations are iterated until \textcolor{black}{tight} convergence is achieved. Finally, to accelerate the convergence of our method we utilize a DIIS implementation \cite{pulay1980convergence} which is turned on after the first few iterations, and whose history is reset when transitioning from phase I to phase II of the iterative scheme.

\subsection{\textcolor{black}{Size Extensivity and Size Intensivity}}
\label{sec:sc}

\textcolor{black}{Size extensivity is a central property of coupled-cluster methods which guarantees that the correlation energy of the multi-electronic system scales linearly with system size in the large-system limit as it should. This property is ensured so long as the working equations contain no unlinked diagrams,\cite{goldstone1957derivation,bartlett1978many} which is famously true of the ground state CC equations in equations \ref{eqn: groundE}-\ref{eqn: groundD}, which contain only connected, and therefore linked, diagrams. While the introduction of the pseudoprojection operator in ESPCC complicates the equations, a diagrammatic analysis reveals that all unlinked diagrams cancel out, leaving only linked diagrams in the working equations. Therefore, ESPCC is a fully size extensive method, in parallel to its ground state counterpart.}

\textcolor{black}{Size intensivity, which ensures that a method's excitation energy on one molecule is unchanged by the addition of other noninteracting molecular fragments, is another important quality to achieve for an excited state method. Because ESPCC's excitation energies are energy differences against traditional ground state CCSD, achieving size intensive excitation energies requires that ESPCC's energy contributions from noninteracting molecular fragments that do not participate in the excitation must be exactly equal to those fragments' CCSD energies. To see that this is indeed the case,} imagine that there are two noninteracting molecular fragments, one of which is in its excited state, molecular fragment $A$, and another which is in its ground state, molecular fragment $B$. Under these circumstances, our wavefunction maintains the usual product factorizability due to the local nature of the \textcolor{black}{pseudo}projector in equation \ref{eqn: projector} (indeed, $\hat{P}$, which we will write here as $\hat{P}_A$, only acts on A): 
\begin{equation}
    \ket{\Psi}=\hat{P}_A e^{\hat{T}_A+\hat{T}_B}\ket{\phi ^* _A}\otimes \ket{\phi _B}=\left( \hat{P}_A e^{\hat{T}_A}\ket{\phi ^* _A} \right) \otimes e^{\hat{T}_B} \ket{\phi _B}.
    \label{eqn:sc-prod-factor}
\end{equation}
Note that this factorization would not work if we employed the intuitive, \textcolor{black}{true}
projector $\hat{P}=1-\ket{\phi ^* _0}\bra{\phi ^* _0}$, as it acts globally on both fragments $A$ and $B$.
Due to the product factorization shown in equation \ref{eqn:sc-prod-factor},
the energy expression in equation \ref{eqn: energy} decomposes into
\begin{align}
    &\frac{\bra{{\psi _{0}}_A}e^{-\hat{T}_A}\hat{H}_A\hat{P}_Ae^{\hat{T}_A}\ket{\phi ^* _A}}{\bra{{\psi _0}_A}e^{-\hat{T}_A}\hat{P}_Ae^{\hat{T}_A}\ket{\phi ^* _A}} + \bra{\phi _B}e^{-\hat{T}_B}\hat{H}_Be^{\hat{T}_B}\ket{\phi _B}\notag\\
    &=E_A+E_B\notag\\
    &=E, \label{eqn: size consistent energy}
\end{align}
where the first term in this equation is the excited state energy for fragment $A$ and the second term is the ground state energy expression for fragment $B$. This energy expression shows great promise of size \textcolor{black}{intensivity}, but in order to guarantee size \textcolor{black}{intensivity} in practice it must be shown that the iterative procedure produces the correct ground state amplitudes for fragment $B$, and thus the correct ground state CCSD energy expression for fragment $B$. It can be shown that in a \textcolor{black}{noninteracting} regime where the $i,h,a,p$ indices in the projection belong to fragment $A$ and the remaining $j,k,b,c$ indices belong to fragment $B$, equations \ref{eqn: double} and \ref{eqn: triple} factorize to produce
\begin{align}
    t_i ^a\bra{{\phi _B} _{j} ^{b}}e^{-\hat{T}_B}\hat{H}_Be^{\hat{T}_B}\ket{\phi _B}&=0, \label{eqn: size consistent singles}\\
    t_h ^p\bra{{\phi _B} _{jk} ^{bc}}e^{-\hat{T}_B}\hat{H}_Be^{\hat{T}_B}\ket{\phi _B}&=0. \label{eqn: size consistent doubles}
\end{align}
Note that, for at least the cluster amplitudes which construct our \textcolor{black}{truncated} ESMF reference, $t_h ^p$/$t_{\bar{h}} ^{\bar{p}}$, the $t_i ^a$ amplitudes cannot be zero. Therefore, to satisfy these amplitude equations, the fragment $B$ portion of the equations must be zero, which is exactly equivalent to the ground state CCSD amplitude equations \ref{eqn: groundS} and \ref{eqn: groundD} for fragment $B$. While this is sufficient for proving size \textcolor{black}{intensivity}, we can make an even stronger claim. By plugging these expressions into the phase I iterative method in equation \ref{eqn: phase I}, we find that the iterative update for cluster amplitudes on fragment $B$ is given by

\begin{equation}
    t^{new} _\nu=t^{old} _\nu -\frac{\bra{{\phi _B} _{\nu}}e^{-\hat{T}_B}\hat{H}_Be^{\hat{T}_B}\ket{\phi _B}}{ \Delta _\nu }, 
\end{equation}

\noindent which is exactly equivalent to the \textcolor{black}{traditional} ground state CCSD iterative update.
In other words, at any iteration during phase I of this method,
the cluster amplitudes that are noninteracting with the excitation will
be exactly equivalent to their counterparts in a ground state CCSD calculation
(see Supporting Information for more details).
We would likewise expect that amplitudes that interact only weakly with the
excitation will behave very similarly as in CCSD,
creating a strong connection between the ground and excited state theories.

\section{Results}

\begin{table*}[tb]
    \setlength{\tabcolsep}{8pt}
    \centering
    \begin{tabular}{lccccccc}
        &$1$&$N$&$N^2$&$N^3$&$N^4$&$N^5$&$N^6$  \\ \hline 
        Number of terms &4&17953&7047&5622&3899&2381&293  \\ 
        Min R &-0.074&0.002&0.888&0.921&0.855&0.875&0.996  \\ 
        Mean R &-.056&0.878&0.977&0.988&0.990&0.993&0.999 \\
    \end{tabular}
    \caption{Asymptotic scaling analysis of all terms${}^{a}$}
    \label{tab:timing}
    \raggedright
    \footnotesize ${}^{a}$Presented are the total number of terms belonging to each asymptotic scaling group along with the minimum and mean correlation coefficients for the linear regression, $R$. The high $R$ values in combination with a manual review of the linear fits on the higher scaling terms confirms that the presented theory asymptotically scales at the same $N^6$ scaling as ground state CCSD.
\end{table*}

\subsection{Computational Details}

To implement this theory, we first derived by hand the tensor contractions that would be necessary for the energy and amplitude equations if we were using a full $\hat{T}_3$ operator containing all triples. We then applied an automatic code generator to split each term involving triples into the various sub-terms corresponding to the $O(N^4)$ subset of the triples that are actually included in our $\hat{T}_{\Tilde{3}}$. This guarantees an $N^4$ memory footprint, and, as discussed in the next section, leads to an implementation with an $N^6$ cost scaling.

For all of the forthcoming results, the EOM-CCSD and $\delta$-CR-EOM-CC(2,3),D\cite{piecuch2006single,piecuch2005renormalized,piecuch2015benchmarking,piecuch2002efficient,kowalski2004new,piecuch2009left,fradelos2011embedding} calculations were performed with GAMESS, \cite{schmidt1993general,barca2020recent} while the CISD and FCI calculations were performed with PySCF.\cite{sun2015libcint,sun2018pyscf,sun2020recent} To obtain the timing and scaling data, calculations were performed on a single core on the same machine to ensure consistency. No frozen core approximation was utilized on any of these methods. For our method, we iterate until the maximum amplitude equation residual was no larger than $10^{-10}$, and for all other methods we utilize the default convergence settings. For the ``ESMF/ESMP2" terms in Figure 1, we run an $N^5$-cost ESMP2 calculation,
\cite{clune2020n5}
extract the energy contribution for these terms,
and add it to our CC energy to form our corrected ESPCC energy.

\subsection{Scaling Analysis \textcolor{black}{and Computational Cost}}

With the inclusion of a subset of triples amplitudes, one might begin to wonder whether the scaling of this method may be worse than its ground state counterpart. However, due to the transformation to the TOP orbital basis and the truncation of the ESMF wavefunction, we find that the asymptotic scaling remains equivalent to CCSD at $O(N^6)$. In order to verify this scaling, we collected a series of timing data in order to analyze the scaling contribution of each term in the current implementation. To collect this data, we timed the contribution from each term in the theory on a system of spatially separated, minimal basis H$_2$ molecules ranging from 5 to 25 molecules. The timing data for each term was then plotted on a log-log plot and fit to a least squares linear regression. The slope resulting from each regression was conservatively rounded up to the nearest whole number and binned according to scaling. These results are reported along with the minimum and mean coefficient of correlation for the linear regression, $R$, in Table \ref{tab:timing}. It is worth noting that the maximum slope for any regression was 5.829, thereby confirming our proposed $O(N^6)$ scaling. Furthermore, each of the linear regressions for the $N^6$ scaling terms was manually reviewed to ensure that the plots did indeed appear well fit by the regression. For a number of the lower scaling terms, the coefficient of correlation indicates a lower quality for the linear regression. Because many of these terms have timings on the order of milliseconds, we suspect that a combination of the overhead for the matrix multiplication functions or the random variations in processor performance may be obfuscating the exact linear relationship. However, we have no reason to believe they would affect the method's asymptotic $N^6$ scaling. Note that, while some optimization in the form of tensor contraction ordering was done to ensure an $O(N^6)$ scaling, no other optimization was performed for this implementation. For example, we have not yet attempted any factorization of the amplitude equations nor have we consolidated the terms which are related through simple permutations of each other, and so the number of terms in Table \ref{tab:timing} is unnecessarily large. With a more careful implementation, we suspect that a large reduction in the total number of terms and wall time will be possible.

\textcolor{black}{At this point, it is worth noting that if one desires to find multiple excited states, for each excited state an independent ESMF starting point would need to be determined followed by an independent ESPCC calculation. While the need for a separate calculation for each different state is typical amongst state-specific CC methods, it does noticeably differ from methods such as EOM-CC which through their diagonalization of the similarity transformed Hamiltonian can determine multiple states simultaneously.}

\subsection{Small basis testing}

For the initial evaluation of ESPCC, we studied its performance on H$_6$ \textcolor{black}{(Table \ref{tab: H6})} and H$_2$O \textcolor{black}{(Table \ref{tab: H2O})} in a 6-31G basis. With these small molecules in a small basis, we are able to compare the errors of this method directly with FCI.
The excitation energy errors with respect to FCI are reported in Figure \ref{fig:excitation}. For H$_6$, we have one H$_2$ in the center with a slightly stretched bond surrounded by two other H$_2$ with equilibrium bond lengths approximately 3\AA { }away. We see that EOM-CCSD and ESPCC both with and without the ESMP2-based correction almost perfectly match both the absolute and excitation FCI energies, while ESMF and ESMP2 have a little more difficulty,
presumably due to their less complete correlation treatments.

For H$_2$O, our method and EOM-CCSD perform comparably for absolute energies, but the sign of the error is very important. The ground state CCSD calculation errors high for the ground state energy, so EOM-CCSD erring low for the excited state energy causes it to perform worse than ESMP2 and both the uncorrected and corrected versions of ESPCC for the excitation energy. This cancellation of errors is important when determining excitation energies, and results in our method performing almost an order of magnitude better than EOM-CCSD in this case. Although further study is needed, it would make sense if the present theory, in which ground and excited state are treated on a much more equal footing, had better error cancellation tendencies.

\subsection{Larger basis testing}

\begin{figure*}[h!]
    \centering
    \includegraphics[width = 0.95\linewidth]{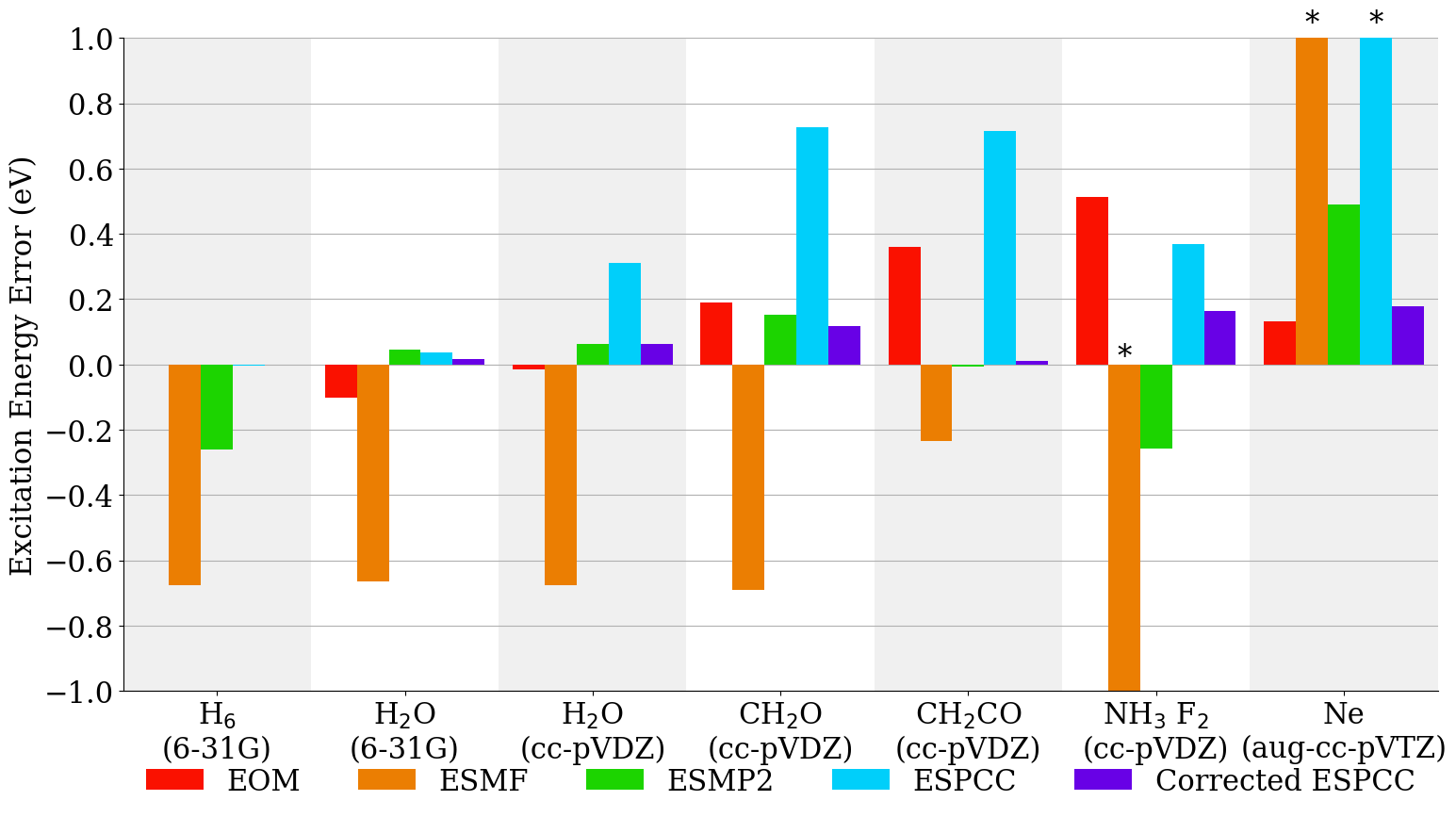}
    \caption{\small The excitation energy errors for (from left to right) EOM-CCSD, ESMF, ESMP2, ESPCC, and Corrected ESPCC on a small set of test molecules. For the two leftmost molecules in the smallest basis, the reference energy was determined via FCI, while for all remaining molecules the reference energy was calculated via $\delta$-CR-EOM-CC(2,3),D. Bars marked with an `*' extend beyond the limits of the plot, precise numbers can be found in the tables above.}
    \label{fig:excitation}
\end{figure*}

With promising results in our initial assessment, we now test these methods in the slightly larger cc-pVDZ basis on the excited states of H$_2$O \textcolor{black}{(Table \ref{tab: H2O})}, formaldehyde (CH$_2$O) \textcolor{black}{(Table \ref{tab: CH2O})}, ketene (CH$_2$CO) \textcolor{black}{(Table \ref{tab: CH2CO})}, and ammonia difluorine (NH$_3$ F$_2$) \textcolor{black}{(Table \ref{tab: NH3F2})}. Of these excitations, the ammonia difluorine charge transfer excitation is especially noteworthy as state-specific orbital relaxations should be particularly important. Finally, we also test these methods on neon \textcolor{black}{(Table \ref{tab: Ne})} in an aug-cc-pVTZ basis in order to investigate performance in a Rydberg excitation. Because both the molecule size and basis size have grown, using FCI as a reference is no longer convenient. Therefore, we have used the triples corrected, $N^7$ scaling CR-CC(2,3) and $\delta$-CR-EOM-CC(2,3),D as our new ground and excited references respectively.
The excitation energies errors with respect to $\delta$-CR-EOM-CC(2,3),D are reported in Figure \ref{fig:excitation}.

\textcolor{black}{In these larger basis sets especially, it is evident that the inclusion of the ESMP2-based correction is critical for maintaining a competitive accuracy for ESPCC. As a result, future work should explore the development of versions of ESPCC which do not need to rely on an ESMP2 correction for missing terms. For now, the ESMP2-based correction provides a reasonable estimate of the influence of these terms, and thus the ESMP2-corrected ESPCC will be the primary focus of the following comparisons with other methods.}

For H$_2$O, EOM-CCSD has both a better absolute and excitation energy as compared to ESPCC. However, in formaldehyde we see the opposite trend, with ESPCC

\clearpage
\onecolumngrid

\begin{table*}[h!]
    \scriptsize
    \centering
    \begin{tabular}{|l | S[table-format=3.7] S[table-format=3.7] | S[table-format=3.4]| }
        \hline
        \multicolumn{1}{|c|}{\multirow{2}{*}{Method}}     & \multicolumn{2}{c|}{{Energy (Ha)}} & {Excitation} \\
                   & {Ground} & {Excited} & {Energy (eV)} \\\hline
        FCI        & -3.437120 & -2.935923 & 13.6383 \\
        ESMF       & -3.356782 & -2.880453 & 12.9616 \\
        ESMP2      & -3.411371 & -2.919699 & 13.3791 \\
        ESPCC      & -3.437119 & -2.935919 & 13.6386 \\
        Corrected ESPCC & -3.437119 & -2.935919 & 13.6383 \\
        EOM-CCSD        & -3.437119 & -2.935910 & 13.6386 \\\hline
    \end{tabular}
    \caption{\small H$_6$ ground, excited, and vertical excitation energies in the 6-31G basis.}
    \label{tab: H6}
\end{table*}

\begin{table*}[h!]
    \scriptsize
    \centering
    \begin{tabular}{|l | S[table-format=4.7] S[table-format=4.7] | S[table-format=2.4]|| S[table-format=4.7] S[table-format=4.7] | S[table-format=2.4]| }
        \hline
        \multicolumn{1}{|c|}{} & \multicolumn{3}{c||}{6-31G Basis} & \multicolumn{3}{c|}{cc-pVDZ Basis} \\\hline
        \multicolumn{1}{|c|}{\multirow{2}{*}{Method}}     & \multicolumn{2}{c|}{{Energy (Ha)}} & {Excitation} & \multicolumn{2}{c|}{{Energy (Ha)}} & {Excitation} \\
                   & {Ground} & {Excited} & {Energy (eV)} & {Ground} & {Excited} & {Energy (eV)} \\\hline
        FCI                        & -76.120137 & -75.803955 & 8.604 & {---} & {---} & {---} \\
        $\delta$-CR-EOM-CC(2,3),D  & {---}      & {---}      & {---} & -76.242856 & -75.937963 & 8.297 \\
        ESMF                       & -75.984322 & -75.692508 & 7.941 & -76.027022 & -75.747005 & 7.620 \\
        ESMP2                      & -76.112169 & -75.794302 & 8.650 & -76.230220 & -75.923046 & 8.359 \\
        ESPCC                      & -76.118633 & -75.801109 & 8.641 & -76.239548 & -75.923188 & 8.609 \\
        Corrected ESPCC            & -76.118633 & -75.801860 & 8.621 & -76.239548 & -75.932350 & 8.359\\
        EOM-CCSD                   & -76.118633 & -75.806267 & 8.501 & -76.239548 & -75.935170 & 8.283\\\hline
    \end{tabular}
    \caption{\small Water (H$_2$O) ground, excited, and vertical excitation energies in the 6-31G and cc-pVDZ bases.}
    \label{tab: H2O}
\end{table*}

\begin{table*}[h!]
    \scriptsize
    \centering
    \begin{tabular}{|l | S[table-format=5.7] S[table-format=5.7] | S[table-format=2.4]| }
        \hline
        \multicolumn{1}{|c|}{\multirow{2}{*}{Method}}     & \multicolumn{2}{c|}{{Energy (Ha)}} & {Excitation} \\
                   & {Ground} & {Excited} & {Energy (eV)} \\\hline
        $\delta$-CR-EOM-CC(2,3),D  & -114.221280 & -114.071520 & 4.075 \\
        ESMF                       & -113.877084 & -113.752709 & 3.384 \\
        ESMP2                      & -114.195345 & -114.040002 & 4.227 \\
        ESPCC                      & -114.211046 & -114.034620 & 4.801 \\
        Corrected ESPCC            & -114.211046 & -114.056952 & 4.193 \\
        EOM-CCSD                   & -114.211046 & -114.054292 & 4.266 \\\hline
    \end{tabular}
    \caption{\small Formaldehyde (CH$_2$O) ground, excited, and vertical excitation energies in the cc-pVDZ basis.}
    \label{tab: CH2O}
\end{table*}

\begin{table*}[h!]
    \scriptsize
    \centering
    \begin{tabular}{|l | S[table-format=5.7] S[table-format=5.7] | S[table-format=2.4]| }
        \hline
        \multicolumn{1}{|c|}{\multirow{2}{*}{Method}}     & \multicolumn{2}{c|}{{Energy (Ha)}} & {Excitation} \\
                   & {Ground} & {Excited} & {Energy (eV)} \\\hline
        $\delta$-CR-EOM-CC(2,3),D  & -152.214501 & -152.075519 & 3.782 \\
        ESMF                       & -151.740890 & -151.610512 & 3.548 \\
        ESMP2                      & -152.181539 & -152.042782 & 3.776 \\
        ESPCC                      & -152.196705 & -152.031461 & 4.497 \\
        Corrected ESPCC            & -152.196705 & -152.057364 & 3.792 \\
        EOM-CCSD                   & -152.196705 & -152.044487 & 4.142 \\\hline
    \end{tabular}
    \caption{\small Ketene (CH$_2$CO) ground, excited, and vertical excitation energies in the cc-pVDZ basis.}
    \label{tab: CH2CO}
\end{table*}

\begin{table*}[h!]
    \scriptsize
    \centering
    \begin{tabular}{|l | S[table-format=5.7] S[table-format=5.7] | S[table-format=2.4]| }
        \hline
        \multicolumn{1}{|c|}{\multirow{2}{*}{Method}}     & \multicolumn{2}{c|}{{Energy (Ha)}} & {Excitation} \\
                   & {Ground} & {Excited} & {Energy (eV)} \\\hline
        $\delta$-CR-EOM-CC(2,3),D  & -255.507183 & -255.166706 & 9.265 \\
        ESMF                       & -254.879868 & -254.593956 & 7.780 \\
        ESMP2                      & -255.468537 & -255.137512 & 9.008 \\
        ESPCC                      & -255.493049 & -255.139057 & 9.633 \\
        Corrected ESPCC            & -255.493049 & -255.146565 & 9.428 \\
        EOM-CCSD                   & -255.493049 & -255.133689 & 9.779 \\\hline
    \end{tabular}
    \caption{\small Ammonia difluorine (NH$_3$ F$_2$) ground, excited, and vertical excitation energies in the cc-pVDZ basis.}
    \label{tab: NH3F2}
\end{table*}

\begin{table*}[h!]
    \scriptsize
    \centering
    \begin{tabular}{|l | S[table-format=5.7] S[table-format=5.7] | S[table-format=3.4]| }
        \hline
        \multicolumn{1}{|c|}{\multirow{2}{*}{Method}}     & \multicolumn{2}{c|}{{Energy (Ha)}} & {Excitation} \\
                   & {Ground} & {Excited} & {Energy (eV)} \\\hline
        $\delta$-CR-EOM-CC(2,3),D  & -128.825887 & -127.094052 & 47.126 \\
        ESMF                       & -128.533273 & -126.754857 & 48.393 \\
        ESMP2                      & -128.819179 & -127.069362 & 47.615 \\
        ESPCC                      & -128.820401 & -127.015682 & 49.109 \\
        Corrected ESPCC            & -128.820401 & -127.082043 & 47.303 \\
        EOM-CCSD                   & -128.820401 & -127.083667 & 47.259 \\\hline
    \end{tabular}
    \caption{\small Neon (Ne) ground, excited, and vertical excitation energies in the aug-cc-pVTZ basis.}
    \label{tab: Ne}
\end{table*}

\begin{table*}[h!]
    \setlength{\tabcolsep}{8pt}
    \centering
    \begin{tabular}{ccccccc}
         He atoms & ESMF & ESMP2 & ESPCC & Corrected ESPCC & EOM-CCSD & CISD \\ \hline
         0 & 7.620 & 8.359 & 8.609 & 8.359 & 8.283 & 11.600 \\ 
         1 & 7.620 & 8.359 & 8.609 & 8.359 & 8.283 & 12.404 \\ 
         2 & 7.620 & 8.359 & 8.609 & 8.359 & 8.283 & 13.198 \\ 
         3 & 7.620 & 8.359 & 8.609 & 8.359 & 8.283 & 13.983 \\ 
         4 & 7.620 & 8.359 & 8.609 & 8.359 & 8.283 & 14.760 \\ 
         5 & 7.620 & 8.359 & 8.609 & 8.359 & 8.283 & 15.528 \\ 
    \end{tabular}
    \caption{Excitation energies (eV) across multiple methods for cc-pVDZ water in the presence of a variable number of noninteracting He atoms${}^a$  }
    \raggedright
    \footnotesize ${}^a$These results corroborate the proof of size \textcolor{black}{intensivity} presented earlier for this method, as well as illustrate how large of an influence a lack of size \textcolor{black}{intensivity} like that in CISD has on the quality of excitation energies.
    \label{tab: size consistent}
\end{table*}

\clearpage
\twocolumngrid

\noindent slightly better than EOM-CCSD for both the absolute and excitation energies. For ketene, we observe the largest discrepancy, where our method results in a significantly lower absolute energy than EOM-CCSD which results in an error just under a hundredth of an eV for ESPCC versus an error on the order of tenths of eV for EOM-CCSD.

In the ammonia difluorine system, we study the lowest lying charge transfer which is from the ammonia lone pair to the difluorine. While methods such as EOM-CCSD have the ability to partially relax the orbitals after an excitation, in situations like this one where the charge density shifts dramatically, this partial relaxation may not be enough and can result in significant errors. In contrast, ESPCC builds a CC ansatz atop an already orbital relaxed, excited-state-specific starting point. This contrast manifests itself in our results, where ESPCC achieves a better absolute energy than EOM-CCSD which results in an excited state energy error a third that of EOM-CCSD.

Finally, we compare these methods in the 2s to 3p Rydberg excitation of Ne. While we may expect EOM-CCSD to face similar difficulties here as in charge transfer, it performs slightly better than ESPCC. It is worth noting that in this particular excited state, the size of the ESMP2 based correction is much larger than any previous molecule ($\sim$2 eV).
It is possible that the error associated with the ESMP2 based correction may be
responsible for the bulk of ESPCC's error here.

\subsection{Size \textcolor{black}{Intensive} Energies}

To test whether this method is size \textcolor{black}{intensive} as we expect, we calculate the excitation energy of H$_2$O in the presence of a variable number of very well separated He atoms. The results for these calculations on a variety of methods are shown in Table \ref{tab: size consistent}.
As we can see, most of the methods presented produce the exact same excitation energy regardless of the number of noninteracting He atoms that are introduced.
The exception is CISD, which has well known size consistency\textcolor{black}{, and consequently size intensivity,} issues. 
\textcolor{black}{Because the ground state energy for ESPCC is calculated using traditional CCSD, the size intensivity of ESPCC implies that our method is correctly reproducing the CCSD ground state energy for the additional He atoms.}
Though one may begin to wonder whether the ESMP2 based correction affects the size \textcolor{black}{intensivity} of our method in any way, it is worth noting that this correction involves orbitals localized only on the fragment that is being excited, and thus does not affect the energy on any other fragment.

\section{Conclusion}

We have presented an approach to excited-state-specific coupled-cluster
theory that applies a simple \textcolor{black}{pseudo}projector to a single-reference
coupled-cluster expansion containing singles, doubles,
and some triples in order to build
a CCSD-like wavefunction atop an excited state mean field starting point.
Like its ground state cousin, this ESPCC approach is \textcolor{black}{size extensive, size intensive},
has an $N^6$ cost, and can be optimized by a DIIS-accelerated
diagonal Jacobian approximation.
For amplitudes that do not interact with the excitation,
the working equations simplify to match those of CCSD,
while amplitudes that are only weakly coupled to the
excitation see slightly modified CCSD-like working equations.
With a correction from excited-state-specific perturbation
theory to account for correlation terms incompatible with
the simple \textcolor{black}{pseudo}projector used in this initial exploration of ESPCC,
excitation energies are within 0.2 eV of high-level benchmarks
in tests that include valence states, a Rydberg state,
and a charge transfer state.
This compares favorably with the performance of
ESMP2 and EOM-CCSD on the same states.

In the future, it will be especially interesting to explore
more general forms for the \textcolor{black}{pseudo}projection operator in order to
both eliminate the reliance on perturbation theory corrections
and admit a more full-fledged ESMF \textcolor{black}{reference}.
Given that many of the doubles amplitudes experience working
equations that are little different than those in CCSD,
it will also be interesting to explore whether, at least
for these amplitudes, straightforward analogues
to ground state perturbative triples corrections can
be formulated.
\textcolor{black}{Furthermore, it will be interesting to evaluate the performance of ESPCC on states involving large doubly excited character, and evaluating whether an augmentation of the cluster operator with selective quadruple excitations will be important for maintaining accuracy.}
Finally, a reimplementation in spatial rather than spin
orbitals that exploits factorization and permutation
symmetry should allow the theory to be tested on
a much wider range of molecules and excited states.

\section{Acknowledgements}

This work was supported by the National Science Foundation's
CAREER program under Award Number 1848012. Calculations were performed 
using the Berkeley Research Computing Savio cluster.
H.T. acknowledges that this 
material is based upon work supported by the National Science Foundation 
Graduate Research Fellowship Program under Grant No. 
DGE 2146752. Any opinions, findings, and conclusions or recommendations 
expressed in this material are those of the authors and do not necessarily 
reflect the views of the National Science Foundation.


\section{References}
\bibliographystyle{achemso}
\bibliography{main}

\clearpage
\onecolumngrid

\section{Supporting Information}
\renewcommand{\thesection}{S\arabic{section}}
\renewcommand{\theequation}{S\arabic{equation}}
\renewcommand{\thefigure}{S\arabic{figure}}
\renewcommand{\thetable}{S\arabic{table}}
\setcounter{section}{0}
\setcounter{figure}{0}
\setcounter{equation}{0}
\setcounter{table}{0}
\section{Relevant Proofs}

\subsection{Intermediate Normalization}

\begin{align}
    \braket{\psi _0 | \Psi}&=\frac{1}{\sqrt{2}}\left( \bra{{\phi ^*}_{h} ^{p}}\hat{P}e^{\hat{T}}\ket{\phi _0 ^*}+\bra{{\phi ^*}_{\bar{h}} ^{\bar{p}}}\hat{P}e^{\hat{T}}\ket{\phi _0 ^*}\right)\\
    &=\frac{1}{\sqrt{2}}\left( \bra{{\phi ^*}_{h} ^{p}}\hat{P}(1+\hat{T}+\dots )\ket{\phi _0 ^*}+\bra{{\phi ^*}_{\bar{h}} ^{\bar{p}}}\hat{P}(1 + \hat{T} + \dots)\ket{\phi _0 ^*}\right)\\
    &=\frac{1}{\sqrt{2}}\left( t _{h} ^{p} + t _{\bar{h}} ^{\bar{p}}\right)=1
\end{align}

\begin{align}
    \bra{\psi _0} e^{-\hat{T}} \ket{\Psi}&=\frac{1}{\sqrt{2}}\left( \bra{{\phi ^*}_{h} ^{p}}e^{-\hat{T}}\hat{P}e^{\hat{T}}\ket{\phi _0 ^*}+\bra{{\phi ^*}_{\bar{h}} ^{\bar{p}}}e^{-\hat{T}}\hat{P}e^{\hat{T}}\ket{\phi _0 ^*}\right)\\
    &=\frac{1}{\sqrt{2}}\left( \bra{{\phi ^*}_{h} ^{p}}(1-\hat{T}+\dots )\hat{P}(1+\hat{T}+\dots )\ket{\phi _0 ^*}\right.  \\ \notag
    &\quad \left. +\bra{{\phi ^*}_{\bar{h}} ^{\bar{p}}}(1-\hat{T}+\dots )\hat{P}(1 + \hat{T} + \dots)\ket{\phi _0 ^*}\right)\\
    &=\frac{1}{\sqrt{2}}\left( \bra{{\phi ^*}_{h} ^{p}}\hat{P}(1+\hat{T}+\dots )\ket{\phi _0 ^*}+\bra{{\phi ^*}_{\bar{h}} ^{\bar{p}}}\hat{P}(1 + \hat{T} + \dots)\ket{\phi _0 ^*}\right. \\ \notag
    &\quad \left. -t_h ^p\bra{{\phi ^*}_{0}}\hat{P}(1+\hat{T}+\dots )\ket{\phi _0 ^*}-t_{\bar{h}} ^{\bar{p}}\bra{{\phi ^*}_{0}}\hat{P}(1+\hat{T}+\dots )\ket{\phi _0 ^*}\right)\\
    &=\frac{1}{\sqrt{2}}\left( t _{h} ^{p} + t _{\bar{h}} ^{\bar{p}}\right)=1
\end{align}

\subsection{Size Intensivity of Energy}

\begin{align}
    &\frac{\bra{\psi _0}e^{-\hat{T}}\hat{H}\hat{P}e^{\hat{T}}\ket{\phi ^* _0}}{\bra{\psi _0}e^{-\hat{T}}\hat{P}e^{\hat{T}}\ket{\phi ^* _0}}
    =\frac{1}{\sqrt{2}}\left( \bra{{\phi ^*} _h ^p}e^{-\hat{T}}\hat{H}\hat{P}e^{\hat{T}}\ket{\phi ^* _0}+\bra{{\phi ^*} _{\bar{h}} ^{\bar{p}}}e^{-\hat{T}}\hat{H}\hat{P}e^{\hat{T}}\ket{\phi ^* _0}\right) \\
    &=\frac{1}{\sqrt{2}} \left[ \left(\bra{{\phi ^* _A}_{h} ^{p}}\otimes \bra{\phi ^* _B}\right) \left( e^{-\hat{T}_A} \otimes e^{-\hat{T}_B} \right) \left(\hat{H}_A + \hat{H}_B\right) \hat{P} _A \left( e^{\hat{T}_A}\otimes e^{\hat{T}_B} \right) \left( \ket{{\phi _A ^*}}\otimes \ket{{\phi _B ^*}}\right) \right. \\ \notag
    &\quad + \left. \left(\bra{{\phi ^* _A}_{\bar{h}} ^{\bar{p}}}\otimes \bra{\phi ^* _B}\right) \left( e^{-\hat{T}_A} \otimes e^{-\hat{T}_B} \right) \left(\hat{H}_A + \hat{H}_B\right) \hat{P} _A \left( e^{\hat{T}_A}\otimes e^{\hat{T}_B} \right) \left( \ket{{\phi _A ^*}}\otimes \ket{{\phi _B ^*}}\right) \right]\\
    &=\frac{1}{\sqrt{2}} \left[ \bra{{\phi ^* _A}_{h} ^{p}} e^{-\hat{T}_A} \hat{H}_A \hat{P} _A e^{\hat{T}_A} \ket{{\phi _A ^*}} \otimes \bra{\phi ^* _B} e^{-\hat{T}_B} e^{\hat{T}_B} \ket{{\phi _B ^*}} \right. \\ \notag
    &\quad +  \bra{{\phi ^* _A}_{h} ^{p}} e^{-\hat{T}_A} \hat{P} _A e^{\hat{T}_A} \ket{{\phi _A ^*}} \otimes \bra{\phi ^* _B} e^{-\hat{T}_B} \hat{H}_B e^{\hat{T}_B} \ket{{\phi _B ^*}} \\ \notag
    &\quad + \bra{{\phi ^* _A}_{\bar{h}} ^{\bar{p}}} e^{-\hat{T}_A} \hat{H}_A \hat{P} _A e^{\hat{T}_A} \ket{{\phi _A ^*}} \otimes \bra{\phi ^* _B} e^{-\hat{T}_B} e^{\hat{T}_B} \ket{{\phi _B ^*}}  \\ \notag
    &\quad + \left. \bra{{\phi ^* _A}_{\bar{h}} ^{\bar{p}}} e^{-\hat{T}_A} \hat{P} _A e^{\hat{T}_A} \ket{{\phi _A ^*}} \otimes \bra{\phi ^* _B} e^{-\hat{T}_B} \hat{H}_B e^{\hat{T}_B} \ket{{\phi _B ^*}} \right] \\ 
    &=\frac{1}{\sqrt{2}} \left[ \bra{{\phi ^* _A}_{h} ^{p}} e^{-\hat{T}_A} \hat{H}_A \hat{P} _A e^{\hat{T}_A} \ket{{\phi _A ^*}} 
     +  \bra{{\phi ^* _A}_{h} ^{p}} e^{-\hat{T}_A} \hat{P} _A e^{\hat{T}_A} \ket{{\phi _A ^*}} \otimes \bra{\phi ^* _B} e^{-\hat{T}_B} \hat{H}_B e^{\hat{T}_B} \ket{{\phi _B ^*}} \right. \\ \notag
    &\quad \left. + \bra{{\phi ^* _A}_{\bar{h}} ^{\bar{p}}} e^{-\hat{T}_A} \hat{H}_A \hat{P} _A e^{\hat{T}_A} \ket{{\phi _A ^*}} 
     +  \bra{{\phi ^* _A}_{\bar{h}} ^{\bar{p}}} e^{-\hat{T}_A} \hat{P} _A e^{\hat{T}_A} \ket{{\phi _A ^*}} \otimes \bra{\phi ^* _B} e^{-\hat{T}_B} \hat{H}_B e^{\hat{T}_B} \ket{{\phi _B ^*}} \right] \\ 
    &=\frac{1}{\sqrt{2}} \left[ \bra{{\phi ^* _A}_{h} ^{p}} e^{-\hat{T}_A} \hat{H}_A \hat{P} _A e^{\hat{T}_A} \ket{{\phi _A ^*}} 
     + \bra{{\phi ^* _A}_{\bar{h}} ^{\bar{p}}} e^{-\hat{T}_A} \hat{H}_A \hat{P} _A e^{\hat{T}_A} \ket{{\phi _A ^*}} \right] \\ \notag
    &\quad + \bra{\phi ^* _B} e^{-\hat{T}_B} \hat{H}_B e^{\hat{T}_B} \ket{{\phi _B ^*}}  \\ 
    &=\bra{{{\psi _0}_A}} e^{-\hat{T}_A} \hat{H}_A \hat{P} _A e^{\hat{T}_A} \ket{{\phi _A ^*}}
    +\bra{\phi ^* _B} e^{-\hat{T}_B} \hat{H}_B e^{\hat{T}_B} \ket{{\phi _B ^*}}=E_A + E_B=E
\end{align}

\subsection{Factorizability of Amplitude Equations}

\begin{align}
    &\bra{{\phi ^*} _{hj} ^{pb}}e^{-\hat{T}}\left(\hat{H}-E\right)\hat{P}e^{\hat{T}}\ket{\phi ^* _0}\\
    &= \left(\bra{{\phi_A ^*} _{h} ^{p}} \otimes \bra{{\phi_B ^*} _{j} ^{b}} \right) \left( e^{-\hat{T}_A} \otimes e^{-\hat{T}_B} \right) \left(\hat{H}_A + \hat{H}_B -E\right)\hat{P}_A \left( e^{\hat{T}_A}\otimes e^{\hat{T}_B} \right) \left( \ket{{\phi _A ^*}}\otimes \ket{{\phi _B ^*}}\right)\\
    &=\bra{{\phi_A ^*} _{h} ^{p}}e^{-\hat{T}_A}\left( \hat{H}_A -E \right) \hat{P}_A  e^{\hat{T}_A} \ket{{\phi _A ^*}} \otimes \bra{{\phi_B ^*} _{j} ^{b}} e^{-\hat{T}_B} e^{\hat{T}_B} \ket{{\phi _B ^*}}\\ \notag
    &\quad +\bra{{\phi_A ^*} _{h} ^{p}}e^{-\hat{T}_A} \hat{P}_A  e^{\hat{T}_A} \ket{{\phi _A ^*}} \otimes \bra{{\phi_B ^*} _{j} ^{b}} e^{-\hat{T}_B} \hat{H} _B e^{\hat{T}_B} \ket{{\phi _B ^*}}\\
    &=\bra{{\phi_A ^*} _{h} ^{p}}e^{-\hat{T}_A} \hat{P}_A  e^{\hat{T}_A} \ket{{\phi _A ^*}} \otimes \bra{{\phi_B ^*} _{j} ^{b}} e^{-\hat{T}_B} \hat{H} _B e^{\hat{T}_B} \ket{{\phi _B ^*}}\\
    &=\bra{{\phi_A ^*} _{h} ^{p}}\left(1-\hat{T}_A + \dots \right) \hat{P}_A  \left(1+\hat{T}_A + \dots \right) \ket{{\phi _A ^*}} \otimes \bra{{\phi_B ^*} _{j} ^{b}} e^{-\hat{T}_B} \hat{H} _B e^{\hat{T}_B} \ket{{\phi _B ^*}}\\
    &=\left[\bra{{\phi_A ^*} _{h} ^{p}} \hat{P}_A  \left(1+\hat{T}_A \right) \ket{{\phi _A ^*}} - t  _{h} ^{p} \bra{{\phi_A ^*}} \hat{P}_A  \left(1+\hat{T}_A \right) \ket{{\phi _A ^*}}\right] \otimes \bra{{\phi_B ^*} _{j} ^{b}} e^{-\hat{T}_B} \hat{H} _B e^{\hat{T}_B} \ket{{\phi _B ^*}}\\
    &=  \left[ t  _{h} ^{p} {\bra{{\phi_A ^*} _{h} ^{p}} \hat{P}_A  \ket{{\phi _A ^*} _{i} ^{a}}} - t  _{h} ^{p} { \bra{{\phi_A ^*}} \hat{P}_A  \ket{{\phi _A ^*}}}\right] \otimes \bra{{\phi_B ^*} _{j} ^{b}} e^{-\hat{T}_B} \hat{H} _B e^{\hat{T}_B} \ket{{\phi _B ^*}}\\
    &= t  _{h} ^{p} \bra{{\phi_B ^*} _{j} ^{b}} e^{-\hat{T}_B} \hat{H} _B e^{\hat{T}_B} \ket{{\phi _B ^*}}
\end{align}

\begin{align}
    &\bra{{\phi ^*} _{hjk} ^{pbc}}e^{-\hat{T}}\left(\hat{H}-E\right)\hat{P}e^{\hat{T}}\ket{\phi ^* _0}\\
    &= \left(\bra{{\phi_A ^*} _{h} ^{p}} \otimes \bra{{\phi_B ^*} _{jk} ^{bc}} \right) \left( e^{-\hat{T}_A} \otimes e^{-\hat{T}_B} \right) \left(\hat{H}_A + \hat{H}_B -E\right)\hat{P}_A \left( e^{\hat{T}_A}\otimes e^{\hat{T}_B} \right) \left( \ket{{\phi _A ^*}}\otimes \ket{{\phi _B ^*}}\right)\\
    &=\bra{{\phi_A ^*} _{h} ^{p}}e^{-\hat{T}_A}\left( \hat{H}_A -E \right) \hat{P}_A  e^{\hat{T}_A} \ket{{\phi _A ^*}} \otimes \bra{{\phi_B ^*} _{jk} ^{bc}} e^{-\hat{T}_B} e^{\hat{T}_B} \ket{{\phi _B ^*}}\\ \notag
    &\quad +\bra{{\phi_A ^*} _{h} ^{p}}e^{-\hat{T}_A} \hat{P}_A  e^{\hat{T}_A} \ket{{\phi _A ^*}} \otimes \bra{{\phi_B ^*} _{jk} ^{bc}} e^{-\hat{T}_B} \hat{H} _B e^{\hat{T}_B} \ket{{\phi _B ^*}}\\
    &=\bra{{\phi_A ^*} _{h} ^{p}}e^{-\hat{T}_A} \hat{P}_A  e^{\hat{T}_A} \ket{{\phi _A ^*}} \otimes \bra{{\phi_B ^*} _{jk} ^{bc}} e^{-\hat{T}_B} \hat{H} _B e^{\hat{T}_B} \ket{{\phi _B ^*}}\\
    &=\bra{{\phi_A ^*} _{h} ^{p}}\left(1-\hat{T}_A + \dots \right) \hat{P}_A  \left(1+\hat{T}_A + \dots \right) \ket{{\phi _A ^*}} \otimes \bra{{\phi_B ^*} _{jk} ^{bc}} e^{-\hat{T}_B} \hat{H} _B e^{\hat{T}_B} \ket{{\phi _B ^*}}\\
    &=\left[\bra{{\phi_A ^*} _{h} ^{p}} \hat{P}_A  \left(1+\hat{T}_A \right) \ket{{\phi _A ^*}} - t  _{h} ^{p} \bra{{\phi_A ^*}} \hat{P}_A  \left(1+\hat{T}_A \right) \ket{{\phi _A ^*}}\right] \otimes \bra{{\phi_B ^*} _{jk} ^{bc}} e^{-\hat{T}_B} \hat{H} _B e^{\hat{T}_B} \ket{{\phi _B ^*}}\\
    &=  \left[ t  _{h} ^{p} {\bra{{\phi_A ^*} _{h} ^{p}} \hat{P}_A  \ket{{\phi _A ^*} _{h} ^{p}}} - t  _{h} ^{p} { \bra{{\phi_A ^*}} \hat{P}_A  \ket{{\phi _A ^*}}}\right] \otimes \bra{{\phi_B ^*} _{jk} ^{bc}} e^{-\hat{T}_B} \hat{H} _B e^{\hat{T}_B} \ket{{\phi _B ^*}}\\
    &= t  _{h} ^{p} \bra{{\phi_B ^*} _{jk} ^{bc}} e^{-\hat{T}_B} \hat{H} _B e^{\hat{T}_B} \ket{{\phi _B ^*}}
\end{align}

\newpage

\section{Molecular Geometries}

All geometries are reported in Bohr.

\noindent \bf{H$_6$}
\begin{verbatim}
H  0.000000000 -0.900000000  0.000000000                                                                           
H  0.000000000  0.900000000  0.000000000                                                                           
H  7.000000000  0.200000000  0.000000000                                                                           
H  8.400000000  0.200000000  0.050000000                                                                          
H -7.000000000 -0.200000000  0.050000000                                                                        
H -8.400000000 -0.200000000  0.000000000 
\end{verbatim}

\noindent \bf{Water}
\begin{verbatim}
O  0.000000000  0.000000000 -0.112851412                                                    
H  0.000000000  1.425999575  0.970077421                                                     
H  0.000000000 -1.425999575  0.970077421 
\end{verbatim}

\noindent \bf{Formaldehyde}
\begin{verbatim}
C  0.000000000  0.000000000 -1.126692899                                                    
O  0.000000000  0.000000000  1.111413933                                                     
H  0.000000000  1.746571764 -2.224070962                                                    
H  0.000000000 -1.746571764 -2.224070962
\end{verbatim}

\noindent \bf{Ketene}
\begin{verbatim}
C  0.000000000  0.000000000 -2.433659698                                                    
C  0.000000000  0.000000000  0.034004982                                                     
O  0.000000000  0.000000000  2.198008243                                                     
H  0.000000000  1.764688153 -3.424475103                                                    
H  0.000000000 -1.764688153 -3.424475103
\end{verbatim}

\noindent \bf{Ammonia} $\rightarrow$ \bf{Difluorine}
\begin{verbatim}
N  0.000000000 -0.443921233 -6.652831844                                                   
H  0.000000000  1.331763699 -7.379272802                                                    
H  1.537789200 -1.331763699 -7.379272802                                                   
H -1.537789200 -1.331763699 -7.379272802                                                  
F  0.000000000 -0.443921233  4.685524903                                                    
F  0.000000000 -0.443921233  7.379272802 
\end{verbatim}

\noindent \bf{Neon}
\begin{verbatim}
Ne  0.000000000  0.000000000  0.000000000
\end{verbatim}

\renewcommand{\thesection}{\Roman{section}}
\setcounter{section}{6}

\end{document}